\pdfoutput=1
\documentclass[12pt]{article}        
	
\usepackage{amsmath}
\usepackage{amssymb}
\usepackage{graphicx}
\usepackage{cite}
\usepackage{url}
\usepackage{euscript}
\usepackage[T1]{fontenc}
\usepackage[utf8]{inputenc}

\newcommand{\veps}{\varepsilon}

\begin{document}
\begin{titlepage}
\begin{flushright} 
{\bf IFJPAN-IV-2012-16}
\end{flushright}
\vspace{15mm}
\begin{center}
{\LARGE\bf
A new Monte Carlo study of evolution equation with coherence}%
\footnote{
This work has been partially supported by grant of 
{\em Narodowe Centrum Bada\'n i Rozwoju} LIDER/02/35/L-2/10/NCBiR/2011
and grant of {\em Narodowe Centrum Nauki} DEC-2011/03/B/ST2/02632.
}
\end{center}

\vspace{5mm}
\begin{center}
  \large{\bf M. Slawinska, S. Jadach and K. Kutak}
\end{center}
\vspace{1mm}
\begin{center}
{\em Institute of Nuclear Physics, Polish Academy of Sciences,\\
    ul. Radzikowskiego 152, 31-342, Krak\'ow, Poland}
\end{center}

\vspace{20mm}
\begin{abstract}
We solve CCFM evolution equation numerically using the 
CohRad program based on Monte Carlo methods.
We discuss the effects of removing soft emissions and  non-Sudakov form factor by
comparing the obtained distributions as functions of accumulated transverse momenta 
or fractions of proton's longitudinal momenta. We also compare the solution of the CCFM with the DGLAP 
equation in the gluonic channel. 
Finally, we analyze the infra-red behaviour of solutions using the so-called
diffusion plots.
\end{abstract}

\vspace{30mm}
\begin{flushleft} 
{\bf IFJPAN-IV-2012-16}
\end{flushleft}

\end{titlepage}

\section{Introduction}
The Large Hadron Collider opened up the possibility to scan parton densities over a wide domain of partons kinematics. This allows for detailed studies of various theoretically interesting and phenomenologically relevant dynamical effects taking place during partons evolution, such as: coherence \cite{Ciafaloni:1987ur}, saturation \cite{Gribov:1984tu} or both \cite{Kutak:2011fu, Kutak:2012yr, Deak:2012mx}. 
In the present work we are in particular interested in the coherence effects in the initial state gluon cascade which is modelled by CCFM ~\cite{Ciafaloni:1987ur, Catani:1989yc} evolution equation.
These effects are taken into account by summing up dominant contribution in angular ordered regions of phase space. 
Therefore, in comparison with DGLAP~\cite{DGLAP} equation, CCFM includes some of the interference effects that are subleading 
from the point of view of approximation to leading-order in the ordering in the hard scales 
$\alpha_S\ln q_T^2/\mu^2$ where $q_T$ is the transverse momentum of a t channel gluon. 
Because of this type of ordering, the CCFM equation is applicable also in the domain of low $x$ and 
can be viewed as a bridge between low $x$ and large $x$ physics.
The CCFM equation, due to the coherence effects included, provides parton distribution function (PDF) 
not only at a given fraction of proton longitudinal momentum 
$x$ and transverse momentum accumulated in the gluonic ladder, but accounts also for additional argument $\overline{p}$, related to the maximal 
angle of gluon emissions. Hence, it allows for matching a PDF with a hard process matrix element given 
the scale of the last emission. The last feature makes it particularly interesting for extextension 
of the BFKL~\cite{BFKL} approach.
The classical CCFM equation is linear and predicts unlimited growth of parton densities 
at small $x$. It can, however, be  extended in order to account for saturation by extending  
it by a non-linear term~\cite{Kutak:2011fu, Kutak:2012qk} 
or impose absorptive boundary conditions~\cite{Avsar:2009pf,Avsar:2010ia}.
The CCFM equation has been already studied theoretically in some limiting cases i.e. 
in the low $x$ limit~\cite{Kwiecinski:1995pu, Kwiecinski:1995pw, Avsar:2009pv, Avsar:2009pf} 
and within  Monte Carlo formulation~\cite{Marchesini:1990zy,Jung:2000hk, Jung:2010si,Chachamis:2011rw}. It has been also used in phenomenological 
applications~\cite{GolecBiernat:1997zs,Bacchetta:2010hh,Deak:2010gk,Deak:2011ga}. 
However, the open issues concerning the CCFM are numerous to mention here just the proper 
form of the initial conditions, the details of violation of unitarity, the role of the soft emissions in the low $x$ limit. Also the efficient algorithm for solving it together with evolution of quark densities is still an open problem.
In the present study we reinvestigate systematically aspects of the CCFM equation using the 
Monte Carlo Markov chain approach focusing on the role played by the non-Sudakov form factor 
and the soft parts of the splitting function. The understanding of the effects coming from both parts of the splitting functions is important for investigations of unitarity violation effects in evolution of 
partons~\cite{Deak:2012mx} since the nonlinearities affect the soft emissions. Such investigation 
is possible since the Monte Carlo numerical integration we use \cite{Jadach:2002ce} allows for 
easy handling of singular integrals and therefore for taking into account effects from the full 
splitting function i.e. soft and hard emissions and corresponding form factors. 
The second reason linked  to the use of Markov chain Monte Carlo is to provide 
a new scheme for performing numerically efficient parton shower based on a 
forward evolution and extend it in the future to account for a nonlinear term allowing for 
saturation of gluons as well as for quarks.\\
The paper is organized as follows. In section 2 we introduce an iterative formulation of the CCFM equation suitable for the Markov Chain implementation. In section 3 we present the Monte Carlo algorithm we apply to solve the CCFM equation using the CohRad program. 
In the section 3 we analyze properties of the CCFM equation i.e. the effect of neglecting the Sudakov 
form factor and soft real emissions, and we compare the CCFM to angular ordered DGLAP cascade. 
Finally we perform the analysis of population of different regions of phase space focusing on the diffusion aspects of the considered equations.

\section{CCFM evolution equation}

\begin{figure}[h!]
\center
\includegraphics[width=5cm]{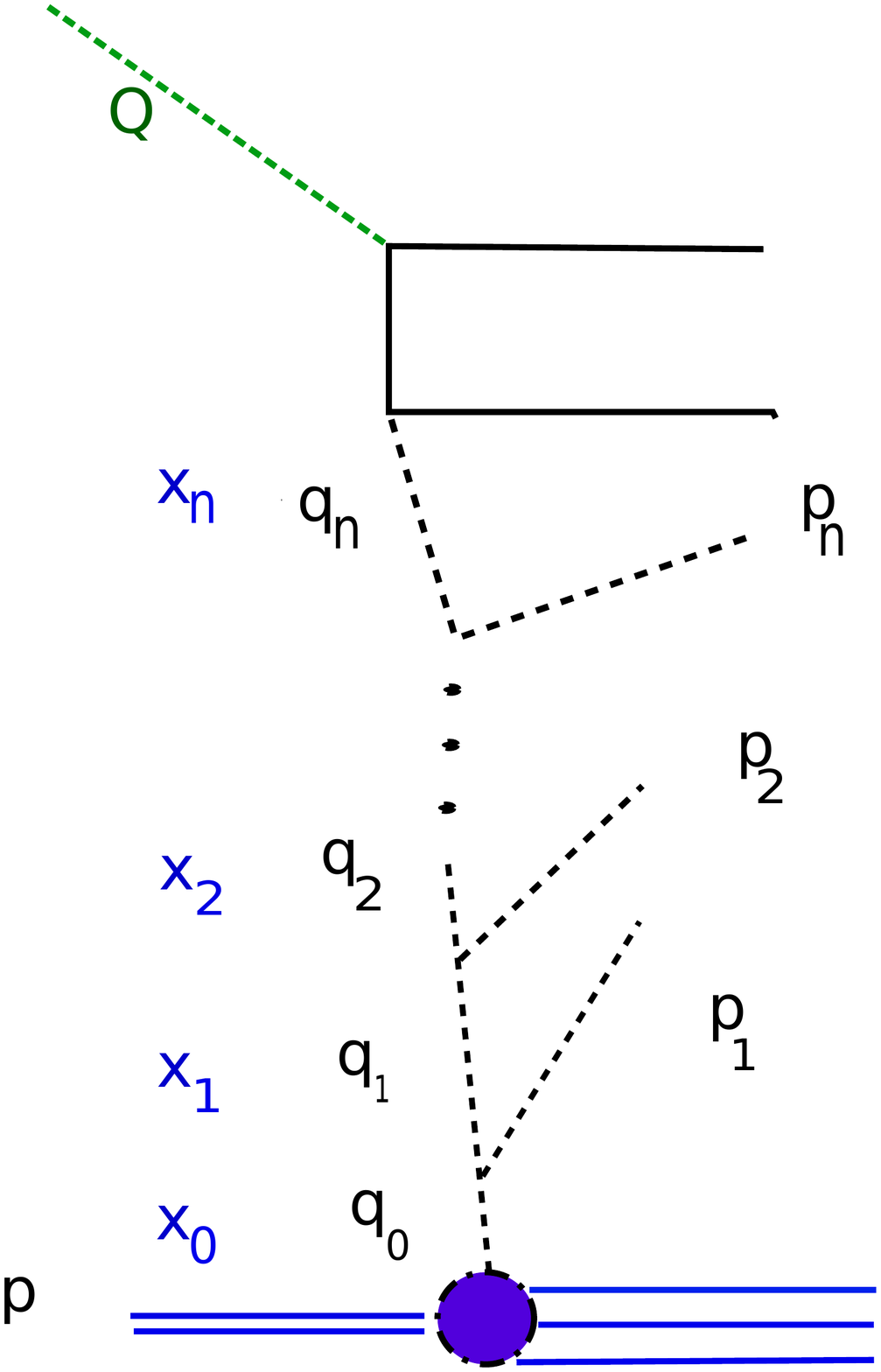}
\caption{Notation used in the text.}\label{fig:notation}
\end{figure}

We use notation as in Fig.~\ref{fig:notation}: $p_i$ and $q_i$ denote four-momenta of real and virtual gluons, respectively, 
and $x_{i+1} = z_{i+1} x_i$ are fractions of longitudinal momenta of the gluon initiating the cascades. 
If $z \simeq 0$, a large momentum fraction has been carried out by a real emitted gluon (hard emission), while $z \simeq 1$ corresponds to a soft emission. 
2-vectors of transverse momenta of the emitted gluons are denoted by ${\bf p}_{i T}$  
and transverse momenta accumulated on the emitting line by 
$ q_{i T} = | {\bf q}_{0 T}  - \sum^i_{j=1} {\bf p}_{j T} |$.  
It is also convenient to use rescaled transverse momenta
$\tilde{\bf p}_i = \frac{{\bf p}_{i T} }{1-z_i}$,
with their modulus being related to angles of emissions:
$\tilde{p}_i = |\tilde{\bf p}_i| =  \frac{p_{i T} }{1-z_i} \simeq E_i \theta_i $.

The CCFM equation imposes angular ordering in the real emissions, that can be expressed 
either using angles: 
$\tilde{p}_{i+1} \geq z_{i}\tilde{p}_i$ or rapidities of the emitted gluons: 
$\ln \frac{p^+_i}{p_{i T}} \equiv \eta_i < \eta_{i+1}$.

In the following we present solution
of the CCFM equation for the unintegrated gluon density in the iterative form:
\begin{equation}
\label{eq:evolution}
\begin{split}
&\mathcal{A}(x, q_T, \bar{p})=
\mathcal{A}(x_0, q_{0_T}, {p}_0)
  \Delta_S(\bar{p}, {p}_0) \delta(x- x_0) \delta(q_T-q_{0_T}) 
\\&~~
+ \sum_{n=1}^{\infty}
  \int d x_0 \int d^2 q_{0T} \;
  \mathcal{A}(x_0, q_{0_T}, {p}_0)
\\&~~~\times
  \int\limits_{\tilde{p}_1<\bar{p}} 
      \frac{d^2 \tilde{p}_1}{\pi \tilde{p}_1^2}
  \int d z_1\;
 \Theta(\tilde{p}_{1} - \tilde{p}_{0})
  \Delta_S(\tilde{p}_{1}, p_0) 
   P_{gg}(z_1, \tilde{p}_1, q_0)
\\&~~~\times 
 \int\limits_{\tilde{p}_2<\bar{p}} 
      \frac{d^2 \tilde{p}_2}{\pi \tilde{p}_2^2} 
 \int d z_2\;
 \Theta(\tilde{p}_{2} - z_1\tilde{p}_{1})
 \Delta_S(\tilde{p}_{2}, z_{1} \tilde{p}_{1}) 
  P_{gg}(z_2, \tilde{p}_2, q_{1T})
\\&~~~
\cdots 
\\&~~~\times 
 \int\limits_{\tilde{p}_n<\bar{p}} 
      \frac{d^2 \tilde{p}_n}{\pi \tilde{p}_n^2} 
 \int d z_n\;
 \Theta(\tilde{p}_{n} - z_{n-1}\tilde{p}_{n-1})
 \Delta_S(\tilde{p}_{n}, z_{n-1} \tilde{p}_{n-1}) 
 P_{gg}(z_i, \tilde{p}_n, q_{n-1,T})
\\&~~~\times 
 \Delta_S(\bar{p}, z_n \tilde{p}_n)\;
 \delta(x - x_0 z_1 \cdots z_n) 
 \delta\big(q_T -|\bf{ q}_{0T}- \sum_{i=1}^n \bf{ p}_{iT}|\big)
\end{split}
\end{equation}
convenient for a Markov chain Monte Carlo implementation. 
The $\Theta(\tilde{p}_{i} - z_{i-1}\tilde{p}_{i-1})$ functions 
impose angular ordering of emissions. 
The scale $\bar{p}$ related to rapidity
position of the hard process will be defined more precisely in the following.
The variable $p_0 =1$ GeV plays the role of 
the minimal scale and the infrared cutoff on transverse momenta, $p_{iT} > p_{0}$.
The initial transverse momentum of a gluon coming from a proton is denoted by $q_{0T}$.
In the above
\begin{equation}
	P_{gg}(z, \tilde{p}, q_T) = \frac{\alpha_S N_C}{\pi} \left( \frac{1}{1-z} + \frac{\Delta_{NS}(\tilde{p}, z, q_T)}{z} \right ) 	
\end{equation}
is the CCFM  splitting function. Its form is similar to the DGLAP~\cite{DGLAP} splitting:
\begin{equation}
	P_{gg}^{DGLAP}(z) = \frac{\alpha_S N_C}{\pi} \left( \frac{1}{1-z} + \frac{1}{z}  + z(1-z) -2 \right  ) 	
\end{equation}
yet does not include terms non-singular in $z\rightarrow 0$. The
non-Sudakov form factor$\Delta_{NS}(p, z, q_T)$ reads:
\begin{equation}
\Delta_{NS}(\tilde{p}, z, q_T) = \exp\left \{ - \frac{\alpha_S N_C}{\pi}
\int_z^1 \frac{dz'}{z'} \int_{z'^2 \tilde{p}^2}^{q_T^2} \frac{d p_T^2}{p_T^2}
 \right \}
 = \exp\left \{ - \frac{\alpha_S N_C}{\pi}
 \ln \frac{1}{z} \ln \frac{q_T^2}{z \tilde{p}^2}
 \right \}.
\end{equation}
The particular form of the non-Sudakov form factor we study here and which allows the form factor 
to be larger than unity was motivated by the investigations in \cite{Avsar:2010ia}. This formulation which we wanted to reproduce in Monte Carlo neglects kinematical constraint effects and leads to fast growth of gluon density towards small values of $q_T$. Studies taking into account kinematical constraint effects we postpone for future investigations. 
In both $\Delta_{ns}$ and $\Delta_s$ as well as in the splitting function we kept $\alpha_s$ constant, $\alpha_s=0.2$, for simplicity at this stage. In the future we plan to use coupling constant as suggested by NLO BFKL results \cite{Kotikov:2000pm}.
This form factor enters the CCFM equation from resummation of virtual emissions that are harder 
than either of the emitting lines. It regulates $z\rightarrow 0$ singularity of the splitting function.  
A detailed discussion on the physical interpretation of the region of integration 
can be found for instance in ref.~\cite{Avsar:2009pf}.

The Sudakov form factor is given by
\begin{equation}\label{eq:sudakov}
\Delta_{S}(q, q')= \exp\left( - \frac{ \alpha_S N_C}{\pi}
		\int^{q^2}_{(q')^2} \frac{d\tilde{p}^2}{\tilde{p}^2} 
		\int_0^{1- \veps(\tilde{p}_i)}\frac{dz}{1-z} \right ).
\end{equation}
It can be interpreted as resummed soft virtual emissions on the emitter line. 
With $\tilde{p}_0$ being the minimal allowed transverse momentum%
\footnote{
 We set $\tilde{p}_0=$1GeV, see below for more discussion.}
the soft (IR) cutoff in CCFM 
$\veps(\tilde{p}_i) = \tilde{p}_0/(\tilde{p}_i x_0)$ 
is evolution scale dependent. 
At the beginning of the evolution  ($\tilde{p}_i \simeq \tilde{p}_0$)
only hard, $z_i\rightarrow 0$, emissions are allowed. 
In DGLAP this cutoff is constant, $\veps(\tilde{p}_i) = \veps$.

Finally let us also mention that the distribution
$\mathcal{A}$ in eq.~\eqref{eq:evolution} is related
to the gluon density $g(x)$ through the relation
\[
xg(x, Q^2) = \int \frac{d^2q_T}{\pi q_T^2 }  x \mathcal{A}(x, q_T, Q^2)\; \Theta(Q^2-q_T^2). 
\]

\section{Monte Carlo algorithm}

We implemented CCFM evolution equation~\eqref{eq:evolution} in CohRad 
as a Markov chain in 
$(\eta_i, x_i)$. Due to the similarities between the structures of CCFM and DGLAP, we applied importance 
sampling and generated emissions from DGLAP distributions that were reweighted to obtain CCFM. 
This is a standard Monte Carlo technique, that improves efficiency of generation.
In the DGLAP evolution the Sudakov form factor has been used as a probability distribution in generating 
each emissions and as a correcting weight for the last emission. The details of the algorithm can be found in ref.~\cite{Jadach:2006ku}.

The emitted particles' four-momenta  are constructed from $\eta_i$ and $x_i$ using the following: 
\begin{equation}\label{eq:mapping}
p_{j_T} = \exp(- \eta_j) \sqrt{s}(x_{j-1} - x_j), \quad
p^+_j=\sqrt{s}(x_{j-1} - x_j), \quad p^-_j = \frac{p_{j_T}^2}{p^+_j},
\end{equation}
It can be checked that $\eta_{j+1} > \eta_j$ hence angular ordering of emissions is imposed by construction.
The maximal allowed rapidity in the hard process frame is equal to 0. Its smallest value is determined by the infrared cutoff on real emissions. 
We set it on the minimal transverse momentum 
$\tilde{p}_0=p_T^{min} = 1GeV$. 
The  rapidity  of the first emitted gluon must therefore obey
\begin{equation}
\tilde{p}_0 \leq e^{-\eta_0}\sqrt{s}x_0,
\end{equation}
which leads to $x_0$-dependent minimum rapidity
\begin{equation}
\eta_0 = \ln \frac{\sqrt{s} x_0}{\tilde{p}_0 }.
\end{equation}

The splitting function used in the Monte Carlo is
\begin{equation}
P_{gg}^{MC} = \frac{\alpha_S N_C}{\pi} \left(\frac{1}{1 - z}+\frac{1}{z} \right  ) .
\end{equation}

The non-Sudakov form factor $\Delta_{NS}(p,z,q)$ is included in the form of a correcting weight for a ``DGLAP-wise''
 generated distributions. 
It multiplies both $1/z$ and $1/(1-z)$ poles.

\section{Results}

We present solutions of CCFM evolution equation for unintegrated gluon distributions 
$x \mathcal{A}(x, q_{T}, \bar{p})$.
They are functions of three parameters:  fractions  of proton's longitudinal momentum $x$, 
transverse momenta $q_T$ accumulated on the emitting line and the hard scale $\bar{p}$.
The role of hard scale $\bar{p}$ in eq.~(\ref{eq:evolution}) 
is to provide an upper limit for $\tilde{p}_i$.

Let us first compare in more detail the implementations of the CCFM  and DGLAP evolution equations,
before moving to analyzing numerical results.
The differences between DGLAP and CCFM lie first of all in the absence of
non-Sudakov formfactor in DGLAP. Both evolution equations include
Sudakov form factor, yet differ in the definition of infrared cutoff
$\veps$, see discussion after equation~\eqref{eq:sudakov}.
The scale-dependent IR cutoff in CCFM suppresses the evolution into low values
of $x$ for small scales, whereas DGLAP allows for emissions in that region.
Another important point is the definition of the hard scale 
$\bar{p}$ in case of CCFM  and DGLAP.
In case of CCFM, following ref.~\cite{Marchesini:1990zy},
it is set to fixed ($x$-independent) angle,
$
\bar{p} = Q.
$
In terms of maximal rapidity, however,
the above leads to $\ln(Q/x)$, see also ref.~\cite{Marchesini:1990zy}.
The variable $x=x_n$ refers to the gluon entering the hard process.
Concerning the hard scale $\tilde{p}_i$, in our implementation of DGLAP
(following ref.~\cite{Jadach:2006ku})
we have the hard scale set as the maximal  
rapidity $\ln(Q)$ and it is  $x$-independent. That translates into
maximal angles, that  are $x$-dependent
$
\bar{p}(x) = xQ.
$
Due to the dependence of $\bar{p}$ on the final $x$, the
eq.~(\ref{eq:evolution}) gets slightly modified:
\begin{equation}
\int\limits_{\tilde{p}_i<\bar{p}} \frac{d^2 \tilde{p}_i}{\pi \tilde{p}_i^2}
\longrightarrow
\int\limits_{\tilde{p}_i<\bar{p}(x_{i-1})} \frac{d^2 \tilde{p}_i}{\pi \tilde{p}_i^2}.
\end{equation}
The above is easily implementable in the Markovian type of MC. 
It is this effect that causes CCFM distribution to rapidly rise at 
small values of $x$.

It is important that for DGLAP the Monte Carlo program enables us to extract 
not only $x$ but also $q_T$ 
and treat the solution of
the DGLAP evolution equation in the following numerical analysis
as a function of three arguments $x$, $q_{T}$ and $\bar{p}$.
The initial distribution will be typically
$x_0 \mathcal{A}_0(x_0, q_{T_0}, \bar{p}) = 1/q_{T_0}^2$
for both, CCFM and DGLAP

\begin{figure}[h!]
\center
\includegraphics[width=79mm, height=54mm]{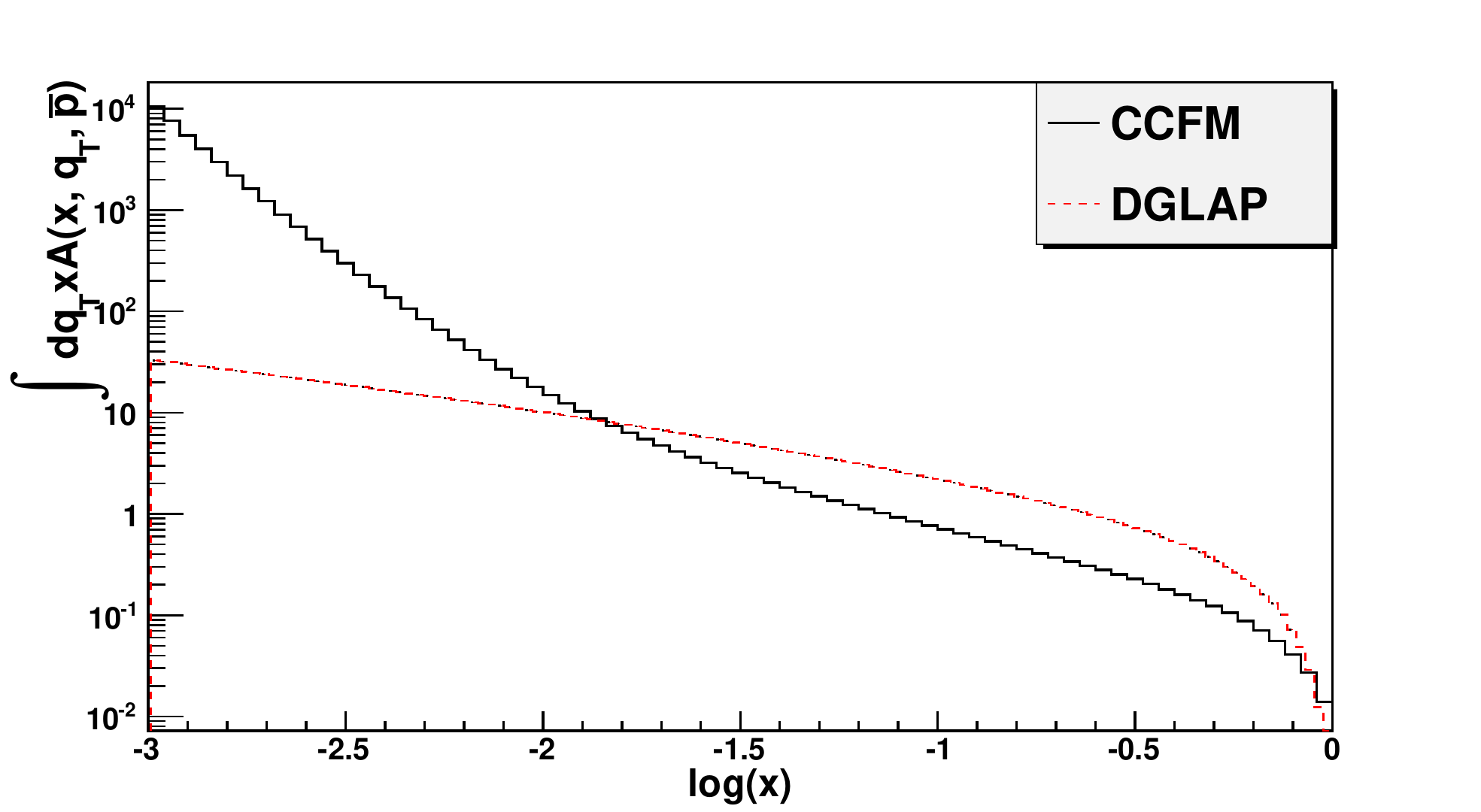}
\includegraphics[width=79mm, height=54mm]{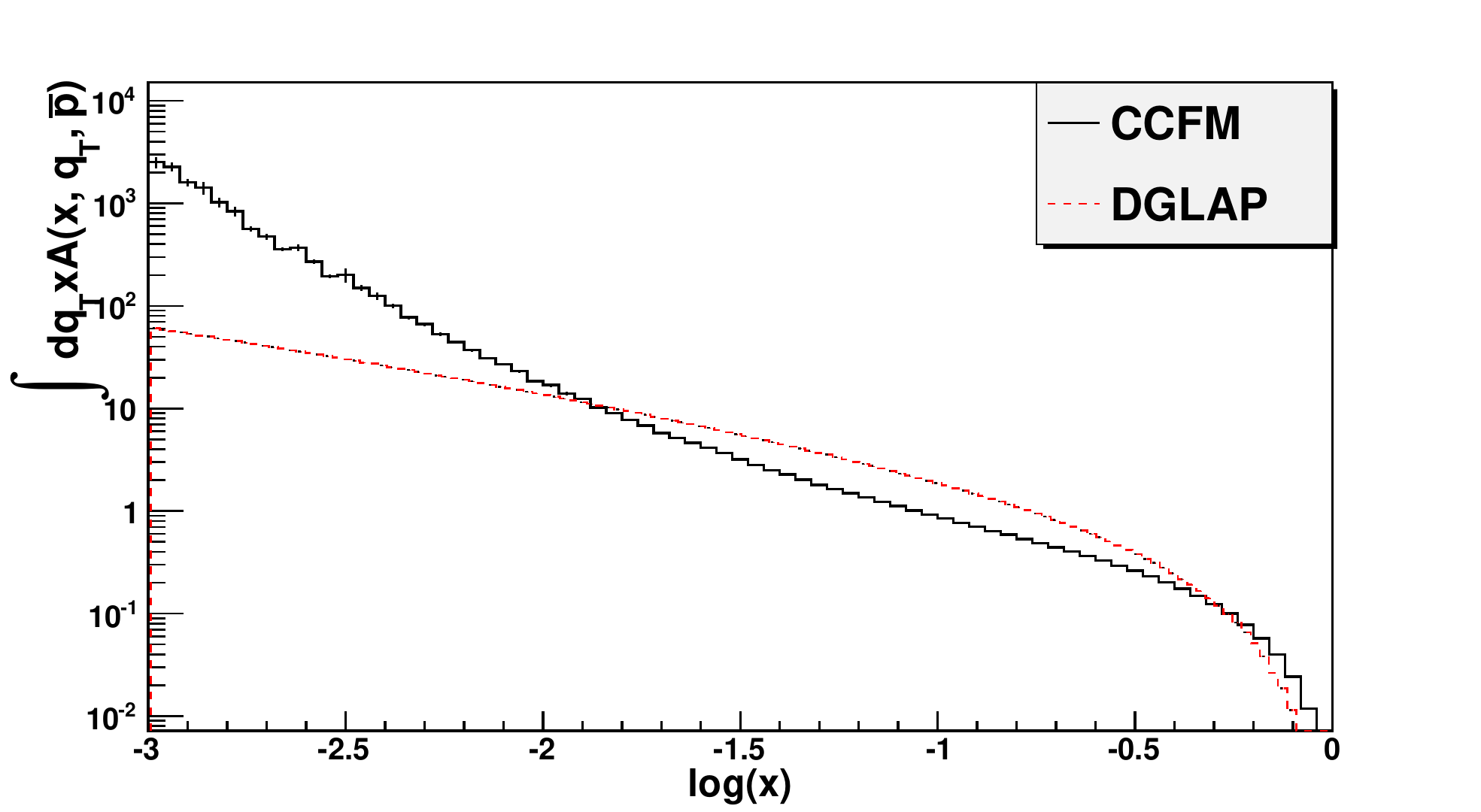}
\caption{
Comparison of distributions in $x$ (integrated over transverse momenta) 
for different hard scales: $Q=$100GeV (left) and 1000GeV (right).}
\label{fig:x}
\end{figure}

In Fig.~\ref{fig:x} the two solutions are shown as functions of $x$ 
(integrated over transverse momenta) at different hard scales $Q$.
In both cases the starting condition is uniform in 
$x$ in order to observe the effect of evolution itself on the distributions. 
It is visible that CCFM equation generates a much sharper rise in the small values of $x$. 
This is caused by the hard scale $\bar{p}$ in CCFM that allows the emitting more gluons,
due to larger (by $\ln(x_0/x)$) allowed rapidity range than in DGLAP.
As a result, more emissions occur, and each of them takes 
away more longitudinal momentum  than in the case of DGLAP.

\begin{figure}[h!]
\center
\includegraphics[width=79mm, height=54mm]{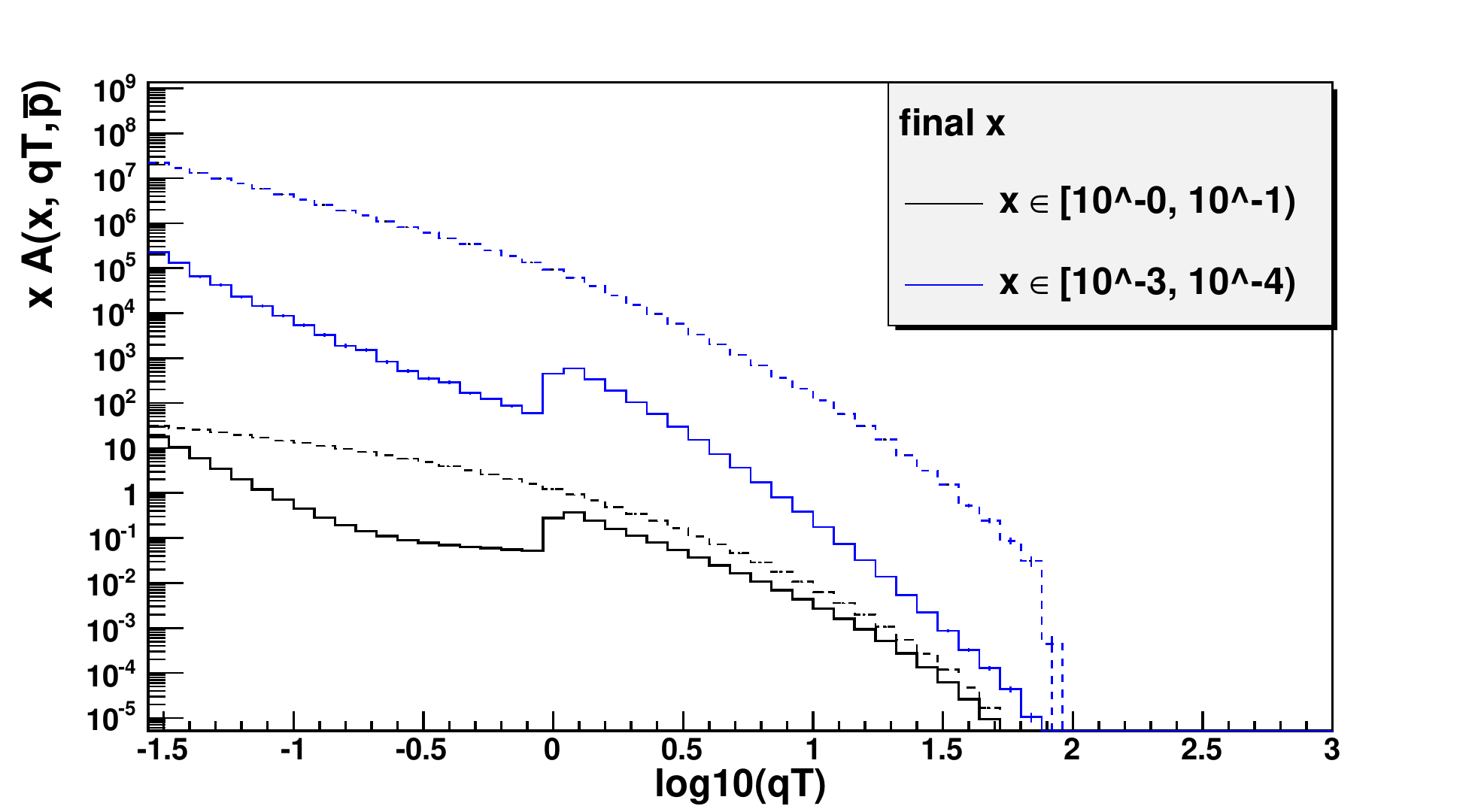}
\includegraphics[width=79mm, height=54mm]{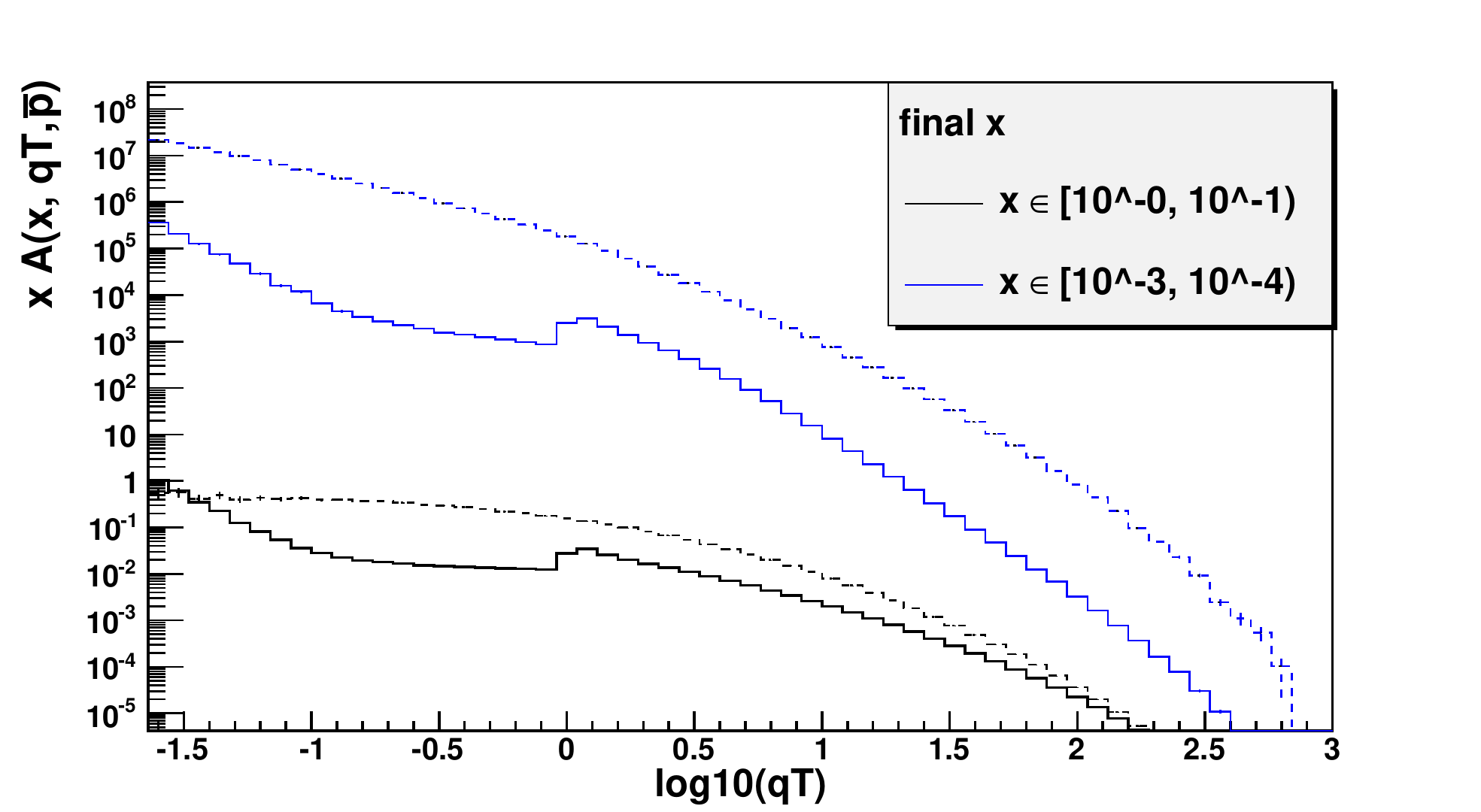}
\caption{
Comparison of solutions obtained in CCFM (solid lines) and DGLAP evolution (dashed lines).
Two plots were made for different maximal angle $Q=$ 100, 1000GeV.
Initial condition: $1/q_T^2 x^0$.
}\label{fig:ccfmvsdglap}
\end{figure}
\begin{figure}[h!]
\center
\includegraphics[width=160mm]{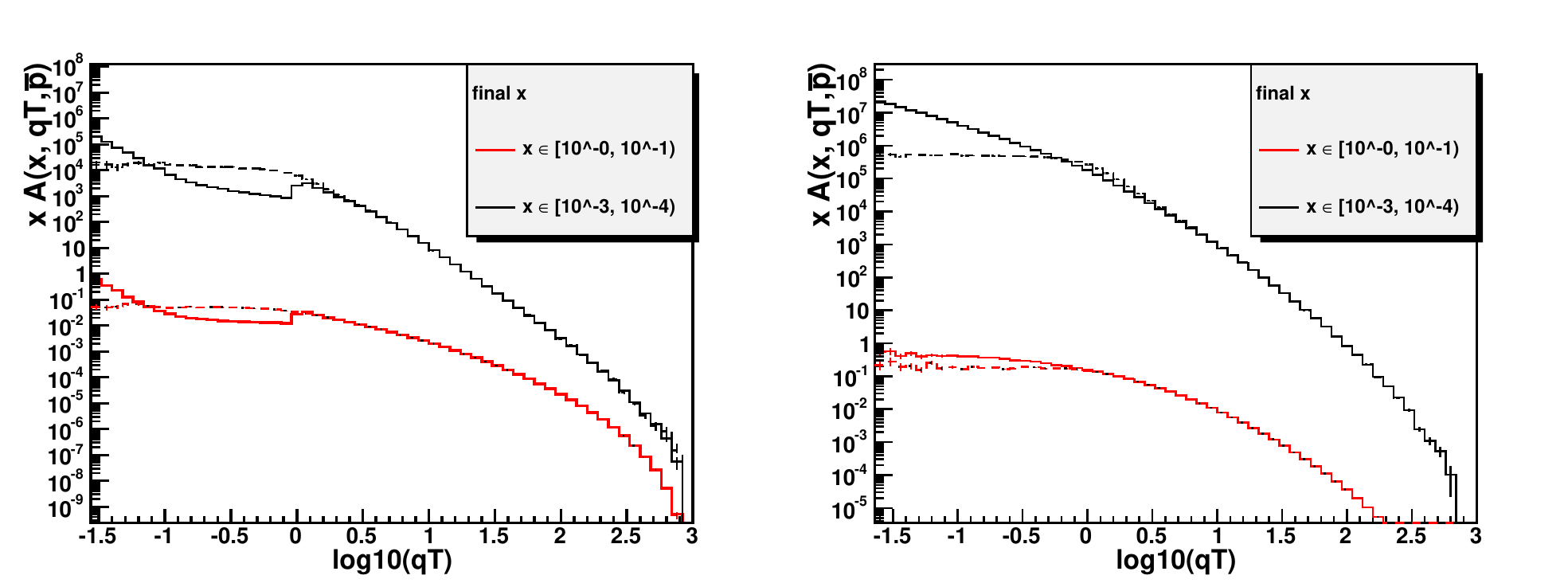}
\caption{Left: CCFM evolution with initial conditions: $1/q_T^2$ (solid lines) and $\exp(-q_T^2)$ (dashed lines).
Right: The same plot for DGLAP solution. 
The hard scale in both plots is $Q=1000$GeV.
}\label{fig:gaussvspow}
\end{figure}

In Fig.~\ref{fig:ccfmvsdglap} we compare CCFM and DGLAP  
distributions of accumulated transverse momenta $q_T$.
Different curves correspond to different values of final $x$, as indicated
in the plots. The relative differences in normalization for different regions in $x$ 
come from the rescaling factors 
$\frac{1}{x \cdot \text{ width of x bin} }$.
In the region below $q_T = 1GeV$ both distributions reflect the initial condition. 
The kink at $q_T =1GeV$ is related to the infrared cutoff on the minimal transverse momenta 
of emitted particles. Above 1GeV the shapes of distributions are generated by evolutions. 
It can be seen that at large values of $x$ both distributions are equivalent. 
At small values of $x$  CCFM distribution  falls down with increasing $q_T$ steeper than DGLAP.
This suppression can be attributed to the non-Sudakov form factor which is smaller than 1
for sufficiently large-angle emissions: $\tilde{p}_{i} > q_{i_T} /\sqrt{z_i}$. 
In order to show that the shape of distributions is influenced by evolution only in the region of large
transverse momenta,
the simulations were repeated with initial conditions $\exp(-q_T^2)$.  
The results are shown in Fig.~\ref{fig:gaussvspow}.

\begin{figure}[h!]
\center
\includegraphics[width=79mm, height=54mm]{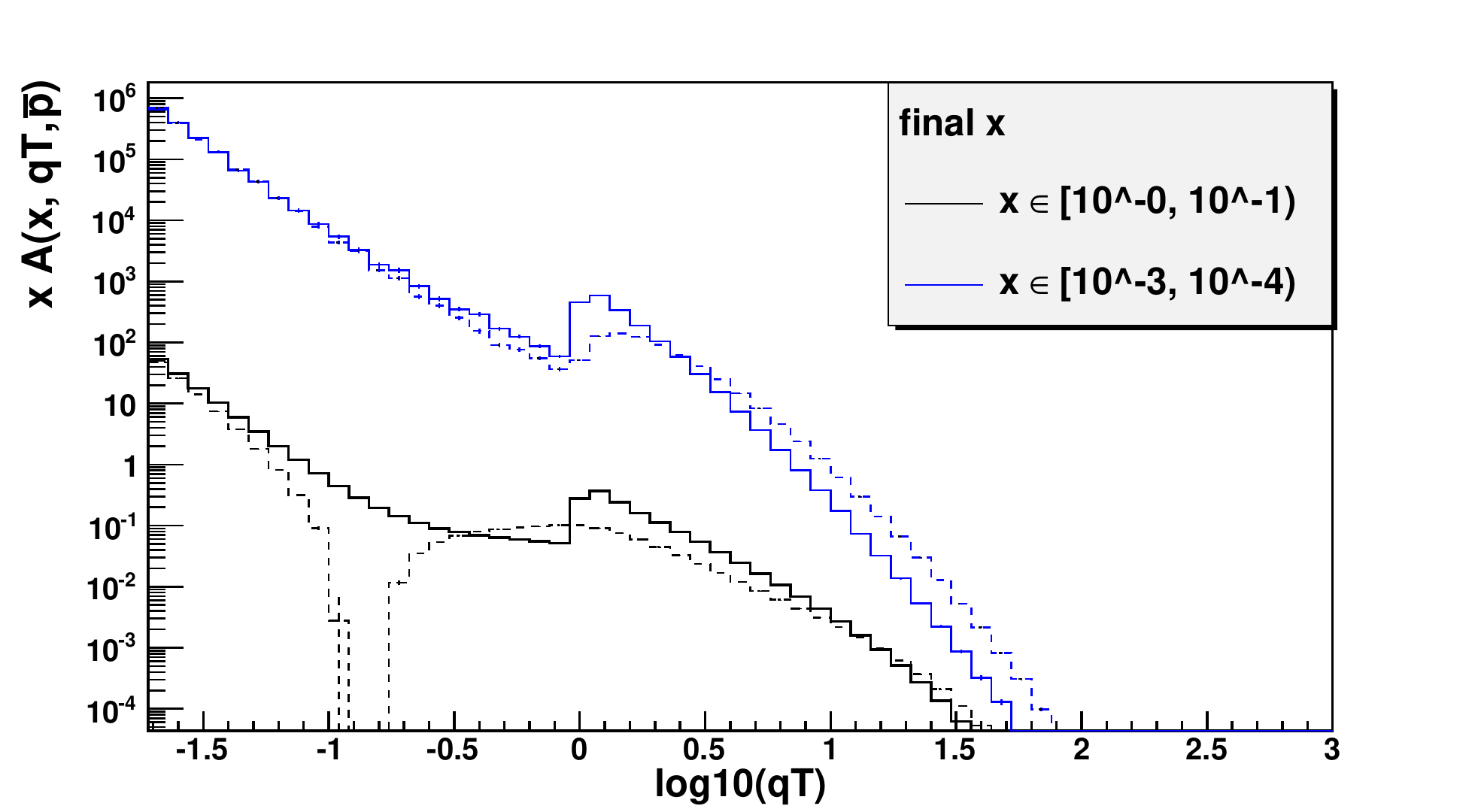}
\includegraphics[width=79mm, height=54mm]{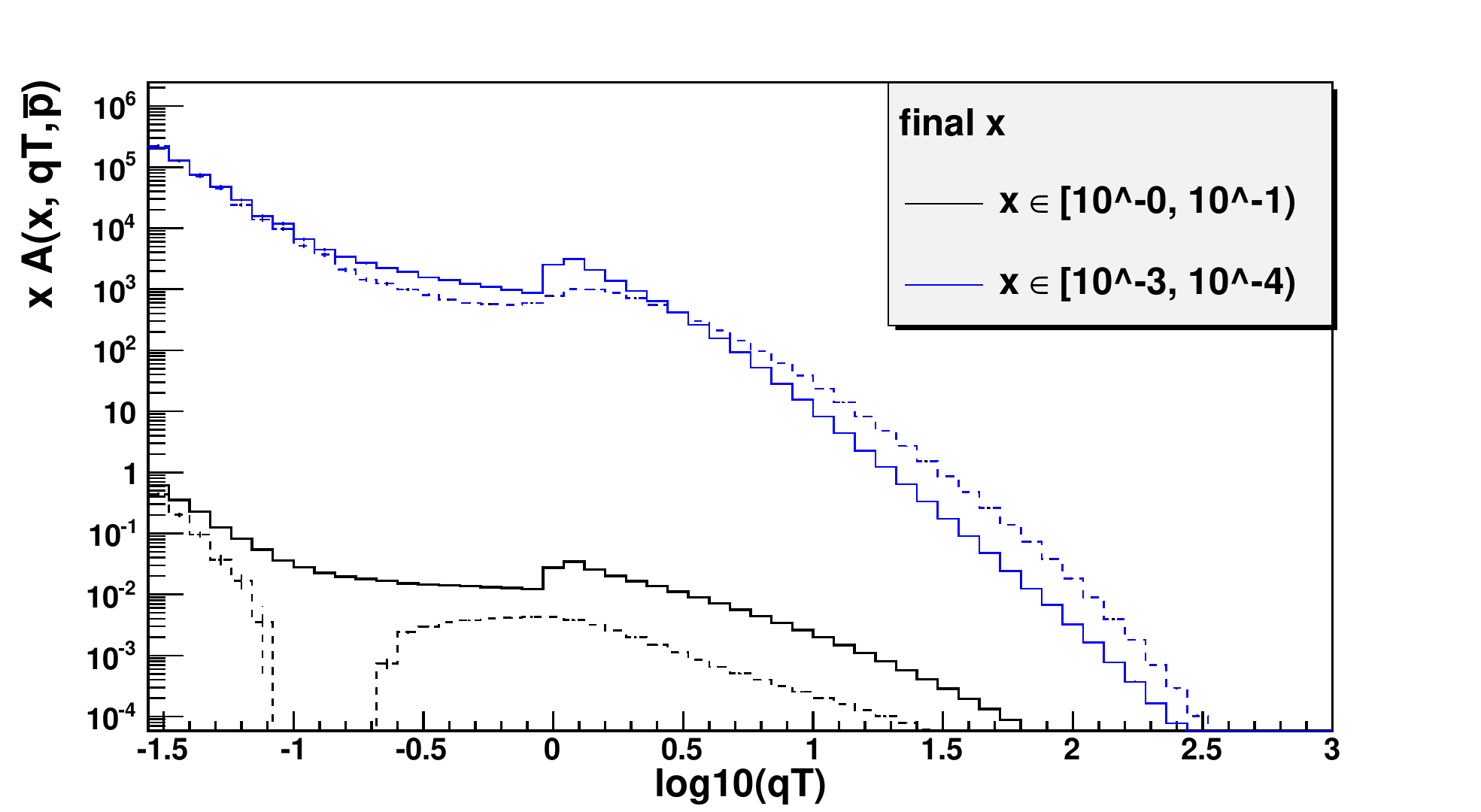}
\caption{Solution of complete CCFM equation (solid lines) and equation with soft emissions removed. 
The hard scale is: 
$Q=$ 100GeV (left), 1000GeV (right). Initial condition  $1/q_T^2$.
}\label{fig:noSudakov}
\end{figure}
The CCFM equation is often solved in a simplified form, with the Sudakov form factor and $1/(1 - z)$ pole 
in the splitting function removed.  
This approximation, described for instance in~\cite{Kwiecinski:1995pu, Avsar:2010ia}, 
is valid in the small $x$ region and for inclusive observables. 
Fig.~\ref{fig:noSudakov} shows comparison between distributions obtained from a complete and simplified equations
for two maximal angles and two values of $x$.
This approximation appears to work well, giving similar slope of the distributions in the region of large transverse momenta
for small $x$ (dark blue lines). The similarity of the distributions at very small values of $q_T$ reflect 
only the common
initial condition. For large values of $x$ the complete and simplified equations diverge, as expected.
Neglection of the soft emissions i.e. Sudakov form factor and $1/(1-z)$ pole manifests itself particularly strongly at low $q_T$ since we observe a sharp cliff on Fig. 5 in small values of $q_T\sim 10^{-1}$.  
In the Monte Carlo simulation this reflects itself as small MC weights given to soft events.

\begin{figure}[h!]
\includegraphics[width=80mm]{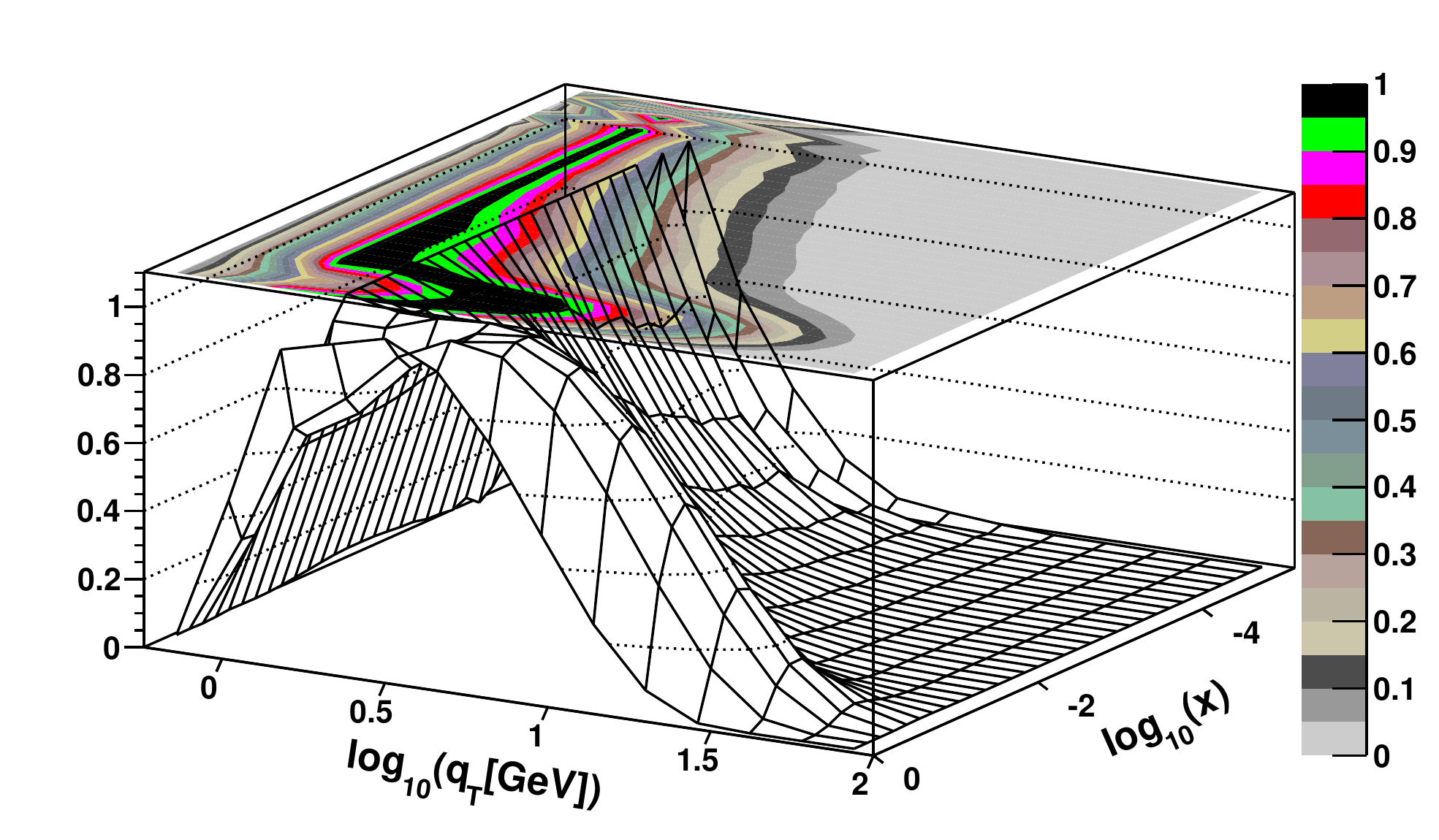}
\includegraphics[width=80mm]{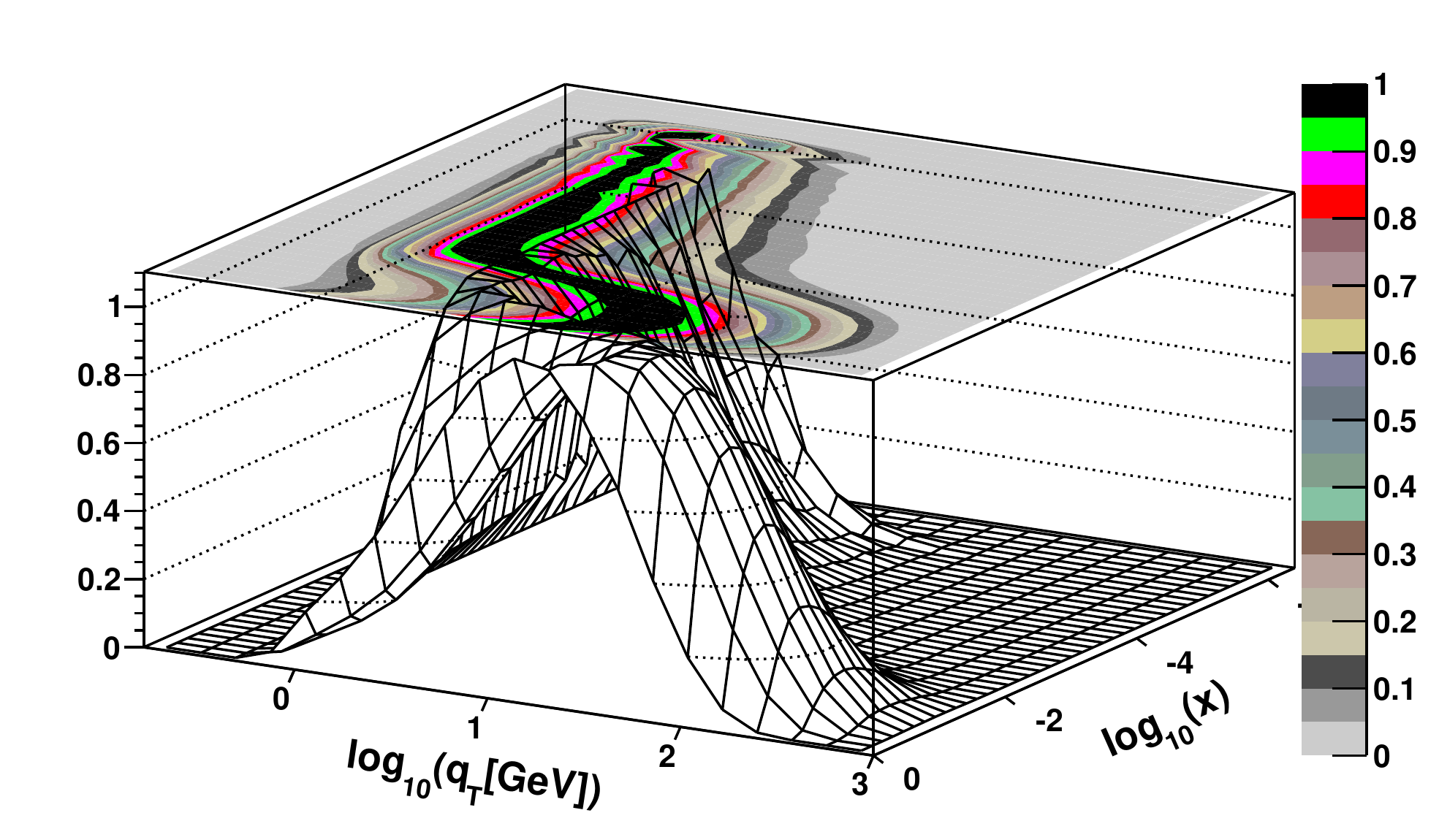}
\caption{The diffusion ratio $\Psi$ of eq.~(\ref{eq:Psi}) for CCFM,
        $Q$ = 100 (left), 1000 (right).}
\label{fig:diffusion_ccfm}
\end{figure}
\begin{figure}[h!]
\includegraphics[width=80mm]{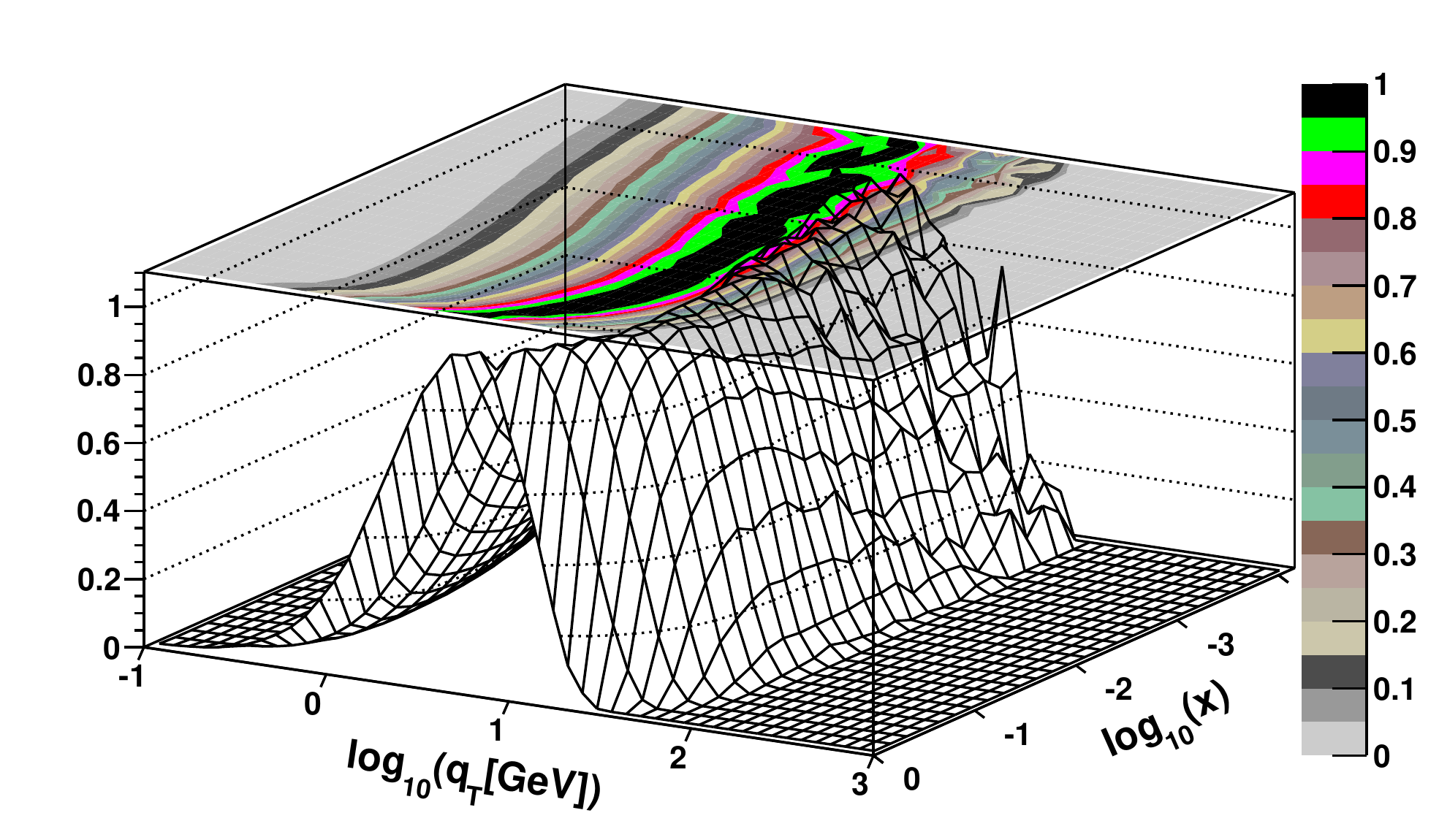}
\includegraphics[width=80mm]{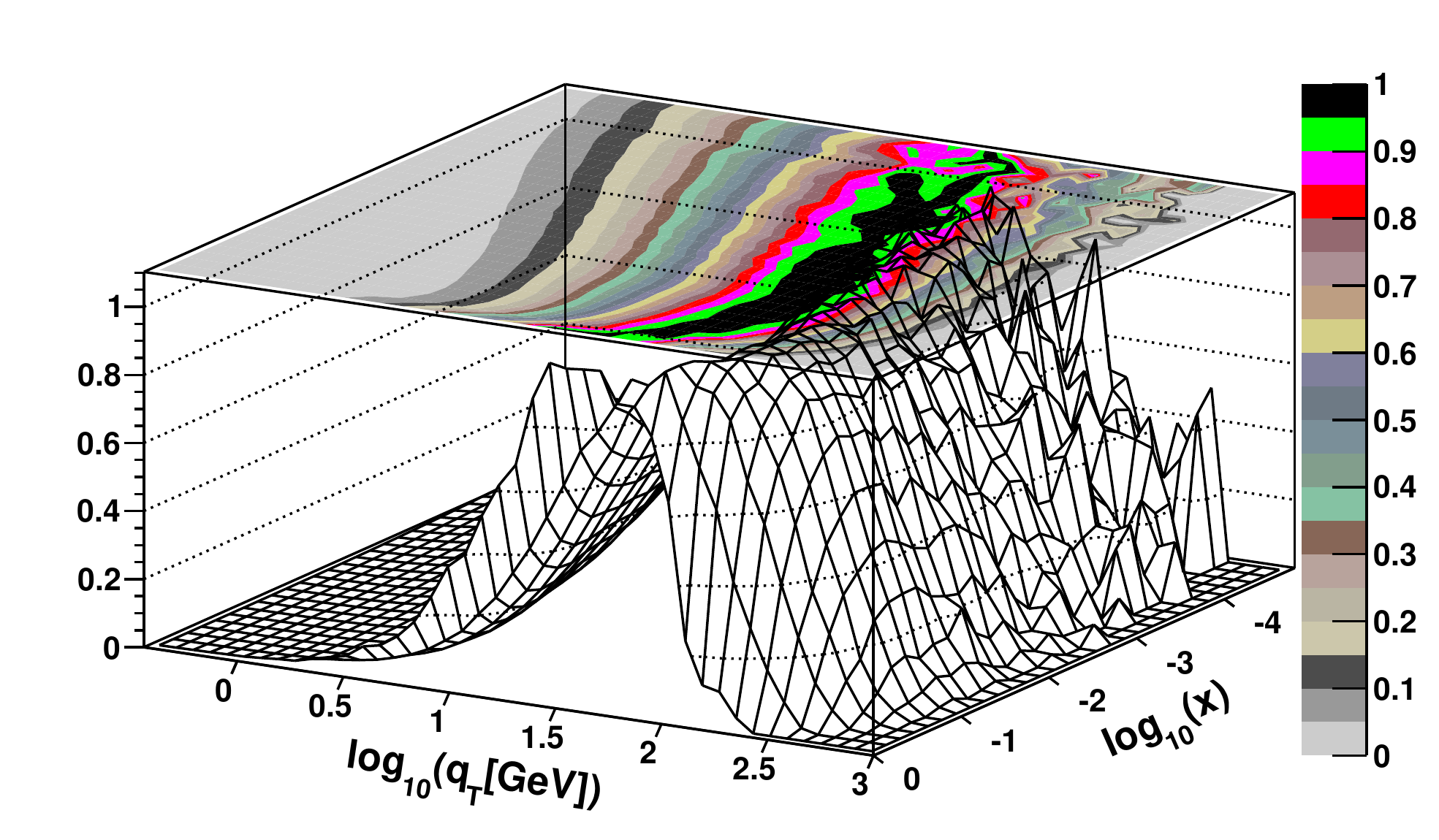}
\caption{The diffusion ratio $\Psi$ of eq.~(\ref{eq:Psi}) for DGLAP.}
\label{fig:diffusion_dglap}
\end{figure}
\begin{figure}[h!]
\includegraphics[width=80mm]{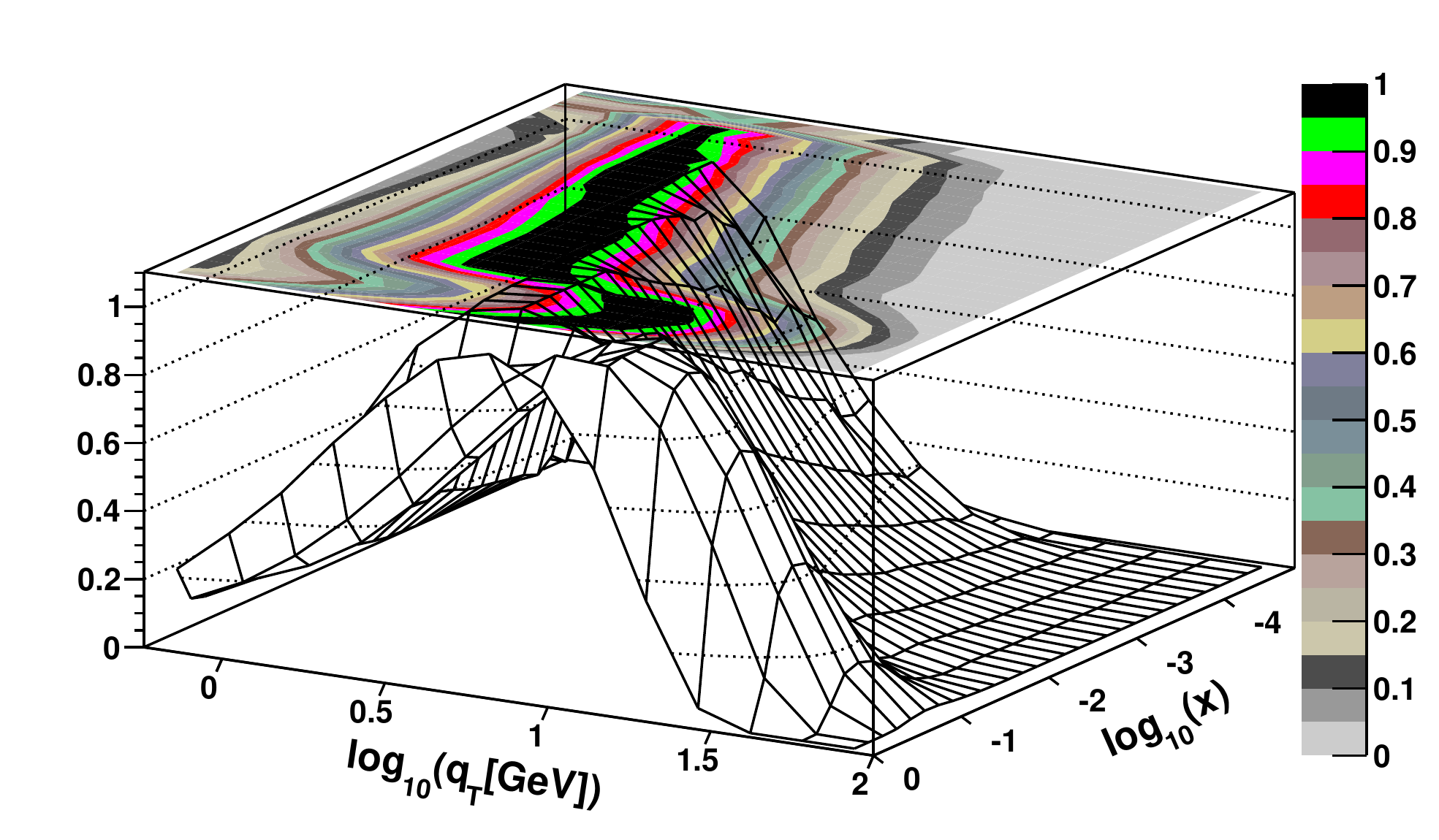}
\includegraphics[width=80mm]{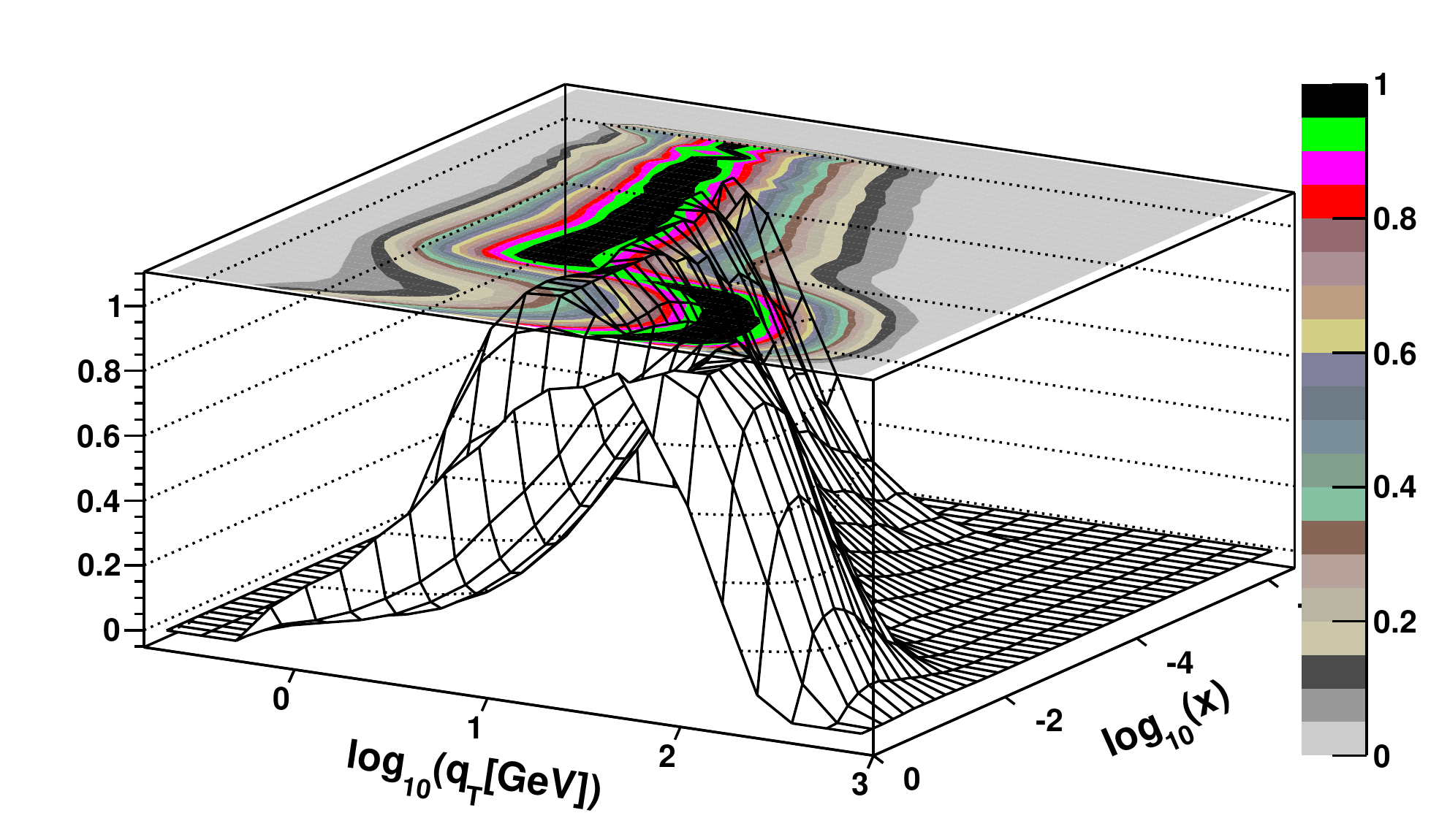}
\caption{The diffusion ratio $\Psi$ of eq.~(\ref{eq:Psi}) for CCFM without soft emissions.}
\label{fig:diffusion_ccfmnoSudakov}
\end{figure}

Another interesting aspect of comparing DGLAP and CCFM equations 
lies in observing the way their solutions populate the  region corresponding to infrared momenta.
In Figs.~\ref{fig:diffusion_ccfm}-\ref{fig:diffusion_ccfmnoSudakov} 
we show two-dimensional distributions $\Psi(x, q_T, \bar{p} )$ obtained from solutions of the evolution equation
$\mathcal{A}(x, q_T,\bar{p})$ 
\begin{equation}
\label{eq:Psi}
\Psi(x, q_T, \bar{p} ) = \frac{q_T x\mathcal{A}(x, q_T,\bar{p}) }{q_T^{max}(x)x\mathcal{A}(x, q_T^{max}(x),\bar{p})  } ,
\end{equation}
where $q_T^{max}(x)$ is the value of transverse momentum, for which $q_T x\mathcal{A}(x, q,_T\bar{p})$
is maximal. In the following, we restrict transverse momenta to the values larger than 1 in order 
to study the effects of evolution only. The form of $\Psi(x, q_T, \bar{p} )$ is motivated by 
the asymptotic behaviour of (linear) BFKL equation, see ref.~\cite{GolecBiernat:2001if} for details.


The diffusion plots in
Figs.~\ref{fig:diffusion_ccfm}-\ref{fig:diffusion_ccfmnoSudakov} are
made according to convention of
ref.~\cite{GolecBiernat:2001if} for all three considered evolutions and hard scales $Q$ = 100, 1000 (left, right plots, respectively).
Fig.~\ref{fig:diffusion_ccfm} shows distribution obtained  from the complete CCFM equation, Fig.~\ref{fig:diffusion_dglap} from DGLAP evolution,
and finally Fig.~\ref{fig:diffusion_ccfmnoSudakov} from CCFM without soft emissions.
On the plots we only show the distribution in the phase space region $q_T > 1GeV$.
Comparing Fig. 6 and Fig. 8 we see that at low $x$ region the solution of the  CCFM is well approximated by just hard emissions. By comparing Fig. 7 to Figs. 6 and 8 we see that the presence of non-Sudakov leads to effectively more diffusive distribution in $q_T$ what is consistent with BFKL like properties of CCFM.

\section{Conclusions}
In this paper we studied some of the properties of the CCFM evolution equation formulated as a 
Markov chain parton shower model. This approach is particularly interesting since it can be 
generalized to a simulation for a nonlinear evolution equation, and will  be presented elsewhere.
The purpose of the paper was to demonstrate capabilities of this method as well as to check 
some of the properties of CCFM which are difficult to quantify by other methods e.g. neglecting contributions on inclusive level of Sudakov form factor and soft emissions at small $x$. 
In the standard non Monte Carlo calculations it is a formidable task to perform a numerical 
calculation with both $1/z$ and $1/(1-z)$ poles included. Our calculation supports the claim 
that at low $x$ one can safely neglect contribution of both Sudakov form factor and 
soft emissions i.e. $1/(1-z)$ pole. Another subject we studied concerned the comparison of the 
CCFM dynamics with DGLAP type of resummation. We observed that the CCFM equation 
due to angular ordering and particular condition for the maximal angle of emitted partons 
generates much faster growth of gluon density than DGLAP at small $x$. 
Finally we addressed the question how the partons generated according to CCFM and DGLAP
equations populate the $(x,q_T)$ plane. The last study will be useful for future investigations of regions of phase space where the unitarity is violated bringing the need for suppression of growth 
of gluons by their recombination.

\section*{Acknowledgements}
We would like to thank Wieslaw P\l{}aczek and Anna Sta\'sto for usefull discussions and Aleksander Kusina for proofreading.
This research has been partially supported by grant of 
{\em Narodowe Centrum Bada\'n i Rozwoju} LIDER/02/35/L-2/10/NCBiR/2011
and grant of {\em Narodowe Centrum Nauki} DEC-2011/03/B/ST2/02632.


\providecommand{\href}[2]{#2}\begingroup\raggedright\endgroup

\end{document}